\documentclass{pasj00}
\draft

\newcommand{\beq}{\begin{eqnarray}}
\newcommand{\eeq}{\end{eqnarray}}
\newcommand\p{\partial}
\newcommand{\OmK}{\Omega_{\rm K}}
\newcommand{\vK}{v_{\rm K}}
\newcommand{\qadv}{q_{\rm adv}}
\newcommand{\qvis}{q_{\rm vis}}
\newcommand{\Qadv}{Q_{\rm adv}}
\newcommand{\Qvis}{Q_{\rm vis}}
\newcommand{\qrad}{q_{\rm rad}}
\newcommand{\fadv}{f_{\rm adv}}
\newcommand{\fpadv}{f'_{\rm adv}}
\newcommand{\Dtheta}{\Delta \theta}

\begin{document}

\SetRunningHead{Gu et al}{Advection-Dominated Accretion Disks}

\title{Advection-Dominated Accretion Disks: \\
Geometrically Slim or Thick?}

\author{Wei-Min \textsc{Gu}, \altaffilmark{1}
Li \textsc{Xue}, \altaffilmark{1}
Tong \textsc{Liu}, \altaffilmark{1,2}
and Ju-Fu \textsc{Lu} \altaffilmark{1}
}

\altaffiltext{1}{Department of Physics
and Institute of Theoretical Physics and Astrophysics,
Xiamen University, Xiamen, Fujian 361005, China}
\email{lujf@xmu.edu.cn}
\altaffiltext{2}{Department of Astronomy, Nanjing University, Nanjing,
Jiangsu 210093, China}

\KeyWords{accretion, accretion disks --- black hole physics
--- hydrodynamics}

\maketitle

\begin{abstract}
We revisit the vertical structure of black hole accretion disks
in spherical coordinates.
By comparing the advective cooling with the viscous heating,
we show that advection-dominated disks are geometrically thick,
i.e., with the half-opening angle $\Delta\theta > 2\pi/5$,
rather than slim as supposed previously in the literature.
\end{abstract}

\section{Introduction}

It was known long since that the very basic assumption of the
Shakura-Sunyaev disk (SSD, Shakura \& Sunyaev 1973), that is,
the geometrical thinness of the disk, $H/R \ll 1$, where $H$ is 
the half thickness of the disk and $R$ is the radius
in cylindrical coordinates, would break down for the inner region
of the disk in some specific situations.
For example, when the mass accretion rate $\dot M$ approaches and
surpasses its critical value corresponding to the Eddington luminosity,
radiation pressure will act to huff the inner region of the disk
in the vertical direction; or when the cooling mechanism is inefficient,
so that the temperature in the disk becomes very high, then gas pressure
will act in a similar way.
In either of these two situations, the inner region of the disk will
get geometrically thick, i.e., with $H/R \sim 1$
(e.g., Frank et al. 2002, p.98).
Based on these understandings, two types of models were proposed
more than twenty years ago, namely the optically thick, radiation 
pressure-supported thick disk
(Abramowicz et al. 1978; Paczy\'{n}ski \& Wiita 1980; Madau 1988)
and the optically thin, ion pressure-supported thick disk
(Rees et al. 1982). To avoid mathematical difficulties,
in these models the disk was assumed to be purely rotating,
i.e., with no mass accretion.
However, the very existence of non-accreting thick disks was
thrown into doubt by the discovery of Papaloizou \& Pringle (1984)
that such disks are dynamically unstable to global
non-axisymmetric modes.
Since the work of Blaes (1987), it had been recognized that
it is accretion, i.e., radial matter motion and energy advection
into the central black hole, that can sufficiently stabilize all modes.
Accordingly, the concept of advection dominance was introduced
and two new types of models were constructed, namely the optically thick,
radiation pressure-supported slim disk (Abramowicz et al. 1988)
and the optically thin, ion pressure-supported, advection-dominated
accretion flow (ADAF, Narayan \& Yi 1994; Abramowicz et al. 1995). 
Both these two types of models are popular nowadays.

Slim disks and ADAFs were supposed to be geometrically slim, i.e.,
with $H/R \lesssim 1$, neither thin nor thick.
The reason for this restriction is the following.
As argued by Abramowicz et al. (1995),
the advection factor $\fadv \equiv \Qadv/\Qvis$,
where $\Qadv$ is the advective cooling rate per unit area and
$\Qvis$ is the viscous heating rate per unit area,
should satisfy the relation
\beq
\fadv \gtrsim \left( \frac{H}{R} \right)^2 \ .
\eeq
Obviously, advection can be important only for disks that are not thin.
But the disk cannot be thick either, because the value of $\fadv$ cannot
exceed 1.

Recently, Gu \& Lu (2007, hereafter GL07) addressed a problem in the
slim disk model of Abramowicz et al. (1988, see also Kato et al. 1998).
In this model, the gravitational potential was approximated in the form
suggested by H\={o}shi (1977), i.e.,
\beq
\psi (R,z) \simeq \psi (R,0) + \frac{1}{2} \OmK^2 z^2 \ ,
\eeq
where $\OmK$ is the Keplerian angular velocity. As shown by GL07,
such an approximation is valid only for geometrically thin disks
with $H/R \lesssim 0.2$, and for a larger thickness it would greatly
magnify the gravitational force in the vertical direction.
Accordingly, the widely adopted relationship $H \OmK /c_s =$ constant
can approximately hold only for thin disks as well.
Since formula (1) was derived by using this relationship,
its validity for thicker disks has not been justified.
GL07 noted that, when the vertical gravitational force is
correctly calculated with the explicit potential $\psi(R,z)$,
``slim" disks are much thicker than previously thought.
However, the work of GL07 was still within the framework of
the slim disk model in some sense. In particular, those authors
did not consider the vertical distribution of velocities, but
instead kept the assumption of vertical hydrostatic equilibrium,
\beq
\frac{1}{\rho} \frac{\p p}{\p z} + \frac{\p \psi}{\p z} = 0 \ ,
\eeq
which is a simplification of the more general vertical momentum equation
\beq
\frac{1}{\rho} \frac{\p p}{\p z} + \frac{\p \psi}{\p z}
+ v_R \frac{\p v_z}{\p R} + v_z \frac{\p v_z}{\p z} = 0
\eeq
(e.g., Abramowicz et al. 1997), where $\rho$ is the mass density,
$p$ is the pressure, and $v_R$ and $v_z$ are the cylindrical radial
and vertical velocities, respectively. While the terms containing
$v_z$ in equation (4) can be reasonably dropped for thin disks
because in this case $v_z$ must be negligibly small, it needs a careful
consideration whether the same can be done for not thin disks
(Abramowicz et al. 1997, also see below in \S 2).

Also regarding to the two main features of advection-dominated
disks, i.e., the advection dominance and the slimness, an important
different approach was made earlier by Narayan \& Yi (1995, hereafter NY95).
NY95 considered rotating spherical
accretion flows ranging from the equatorial plane to the rotation axis,
i.e., with $H/R \to \infty$ and with no free surfaces.
They assumed self-similarity in the radial direction and solved
differential equations describing the vertical structure of the flow,
and showed that, comparing to their exact solutions,
the solutions obtained previously with the vertical integration
approach are very good approximations, provided ``vertical" means
the spherical polar angle $\theta$, rather than the cylindrical height $z$.
This seemed to indicate that advection-dominated disks
are not necessarily limited to be slim.
However, those authors did not calculate the advection
factor $\fpadv$ (they defined $\fpadv \equiv \qadv/\qvis$,
with $\qadv$ and $\qvis$ being the advective cooling rate
and the viscous heating rate per unit volume, respectively),
but rather set it a priori to be a constant.
It is still not answered how their $\fpadv$ varies with $\theta$,
or how $\fadv$ per unit area varies with the thickness of the disk,
and what is required for advection to be dominant.

In this work we try to make some complementarity to NY95 and
some refinements to GL07. We consider the vertical structure of
accretion flows with free surfaces and show that advection-dominated
disks must be geometrically thick rather than slim. Our results may
suggest to recall the historical thick disk models mentioned above,
but with improvements that they have to include accretion now.

\section{Equations}

We consider a steady state axisymmetric accretion flow
in spherical coordinates ($r$, $\theta$, $\phi$)
and use the Newtonian potential $\psi = - GM/r$ since it is convenient
for the self-similar formalization adopted below,
where $M$ is the black hole mass. The basic equations of 
continuity and momenta are
\beq
\frac{1}{r^2} \frac{\p}{\p r} (r^2 \rho v_r)
+ \frac{1}{r \sin \theta} \frac{\p}{\p \theta}
(\sin \theta \rho v_{\theta}) = 0 \ , \\
v_r \frac {\p v_r}{\p r} + \frac{v_{\theta}}{r} \left( \frac{\p
v_r}{\p \theta} - v_{\theta} \right) -\frac{v_{\phi}^2}{r} = -
\frac{GM}{r^2} - \frac{1}{\rho} \frac{\p p}{\p r} \ , \\
v_r \frac{\p v_{\theta}}{\p r} + \frac{v_{\theta}}{r} \left(
\frac{\p v_{\theta}}{\p \theta} + v_r \right) -
\frac{v_{\phi}^2}{r} \cot \theta = - \frac{1}{\rho r} \frac{\p
p}{\p \theta} \ , \\
v_r \frac{\p v_{\phi}}{\p r} + \frac{v_{\theta}}{r} \frac{\p
v_{\phi}}{\p \theta} + \frac{v_{\phi}}{r} (v_r + v_{\theta}
\cot \theta ) = \frac{1}{\rho r^3} \frac{\p}{\p r} (r^3 t_{r\phi})
\eeq
(e.g., Xue \& Wang 2005), where $v_r$, $v_{\theta}$, and $v_{\phi}$
are the three velocity components. We assume that only the
$r\phi$-component of the viscous stress tensor is important,
which is $t_{r\phi} = \nu \rho r \p (v_{\phi}/r) /\p r$,
where $\nu = \alpha c_s^2 r / \vK$ is the kinematic viscosity coefficient,
$\alpha$ is the constant viscosity parameter, $c_s$ is the sound speed
defined as $c_s^2 = p/\rho$, and $\vK = (GM/r)^{1/2}$ is the
Keplerian velocity.

We do not simply assume vertical hydrostatic equilibrium
(eq. [3]). Equation (7) is the general vertical momentum equation
in spherical coordinates, corresponding to equation (4)
in cylindrical coordinates. Abramowicz et al. (1997) have given
several reasons why spherical coordinates are a much better choice.
We only mention one of these reasons that is particularly important
for our study here. The stationary accretion disks calculated
in realistic two-dimensional (2D) and three-dimensional (3D)
simulations resemble
quasi-spherical flows, i.e., in spherical coordinates
the half-opening angle of the flow $\Dtheta \approx$ constant,
or in cylindrical coordinates the relative thickness
$H/R \approx$ constant, much more than quasi-horizontal flows,
i.e., $H \approx$ constant (e.g., Papaloizou \& Szuszkiewicz 1994; NY95).
If no outflow production from the surface of the disk is assumed,
then obviously $v_{\theta} = 0$ is a reasonable approximation
for disks with any thickness (Xue \& Wang 2005); but $v_z$ cannot
be neglected for not thin disks because there is a relation
$v_z / v_R \sim H/R$ for quasi-spherical flows, making equation (4)
difficult to deal with.

Similar to NY95, we assume self-similarity in the radial direction
\beq
v_r \propto r^{-1/2}; \ v_{\theta} = 0; \ v_{\phi}
\propto r^{-1/2}; \nonumber \\ 
\rho \propto r^{-3/2}; \ c_s \propto r^{-1/2}. \nonumber
\eeq
The above relation automatically satisfies the
continuity equation (5). By substituting the relation,
the momentum equations (6-8) are reduced to be
\beq
\frac{1}{2} v_r^2 + \frac{5}{2} c_s^2 + v_{\phi}^2 - \vK^2 = 0 \ , \\
\frac{c_s^2}{p} \frac {d p}{d \theta} = v_{\phi}^2 \cot \theta \ , \\
v_r = - \frac{3}{2} \frac{\alpha c_s^2}{\vK} \ .
\eeq
Four unknown quantities, namely $v_r$, $v_{\phi}$, $c_s$ and $p$,
appear in these three equations.
This is because we do not write the energy equation,
whose general form is $\qvis = \qadv + \qrad$, where $\qrad$ is
the radiative cooling rate per unit volume.
In principle, the general energy equation should be solved,
and then $\fpadv$ is obtained as a variable, as done, e.g., by
Manmoto et al. (1997) for ADAFs and by Abramowicz et al. (1988)
and Watarai et al. (2000) for slim disks. But due
to complications in calculating the radiation processes,
in NY95 and even in works on global ADAF solutions
(e.g., Narayan et al. 1997), $\qadv = \fpadv \qvis$ or
$\Qadv = \fadv \Qvis$ was used instead as an energy equation
and $\fpadv$ or $\fadv$ was given as a constant.
Since our purpose here is to investigate the variation of $\fadv$
with the thickness of the disk, we wish to calculate $\Qadv$
and $\Qvis$ respectively, and then estimate $\fadv$.
To do this, we further assume a polytropic relation,
$p = K \rho ^{\gamma}$, in the vertical direction, which is 
often adopted in the vertically integrated models of
geometrically slim disks (e.g., Kato et al. 1998, p.241).
We admit that the polytropic assumption is a simple way
to close the system, and then enables us to calculate the
dynamical quantities and evaluate $\fadv$ self-consistently.

With the polytropic relation
and the definition of the sound speed $c_s^2 = p/\rho$,
equation (10) becomes
\beq
\frac{d c_s^2}{d \theta} =
\frac{\gamma -1}{\gamma} v_{\phi}^2 \cot \theta \ ,
\eeq
which along with equations (9) and (11) can be solved for
$v_r$, $v_{\phi}$, and $c_s$.
A boundary condition is required for solving the differential
equation (12), which is set to be $c_s = 0$
(accordingly $\rho = 0$ and $p = 0$) at the surface of the disk.
The quantities
$\qadv = p v_r (\p\ln p/\p r - \gamma \p\ln \rho/\p r)/(\gamma-1)$
and
$\qvis = \nu \rho r^2 [\p (v_{\phi}/r)/\p r]^2$
are expressed in the self-similar formalism as
\beq
\qadv = - \frac{5-3\gamma}{2(\gamma -1)} \frac{p v_r}{r} \ , \\
\qvis = \frac{9}{4} \frac{\alpha p v_{\phi}^2}{r \vK} \ ,
\eeq
then $\Qadv$ and $\Qvis$ are given by the vertical integration,
\beq
\Qadv = \int_{\frac{\pi}{2}-\Dtheta}^{\frac{\pi}{2}+\Dtheta}
\qadv \ r \sin\theta \ d\theta \ , \\
\Qvis = \int_{\frac{\pi}{2}-\Dtheta}^{\frac{\pi}{2}+\Dtheta}
\qvis \ r \sin\theta \ d\theta \ ,
\eeq
and $\fadv \equiv \Qadv/\Qvis$ is obtained.
In our calculations $\alpha = 0.1$ is fixed.

\section{Numerical results}

We first study the variation of dynamical quantities with the polar angle
$\theta$ for a given disk's half-opening angle $\Dtheta$.
Figure 1 shows the profiles of $v_r$ (the dashed line), $v_{\phi}$
(the dot-dashed line), $c_s$ (the solid line), and $\rho$ (the dotted line)
for three pairs of parameters, i.e.,
$\gamma = 4/3$ and $\Dtheta = 0.25\pi$ for Fig.~1$a$,
$\gamma = 4/3$ and $\Dtheta = 0.45\pi$ for Fig.~1$b$,
and $\gamma = 1.65$ and $\Dtheta = 0.498\pi$ for Fig.~1$c$.
The parameters are marked in Figure~3 by filled stars,
which clearly show the corresponding values of the advection factor $\fadv$.
Obviously, advection is not significant for case $a$ ($\fadv < 0.1$),
but is dominant for cases $b$ and $c$ ($0.5 < \fadv < 1$).
Comparing our results with Fig.~1 of NY95, it is seen that the profiles
of $v_r$ and $\rho$ are similar, i.e., $v_r$ (the absolute value) and $\rho$
increase with increasing $\theta$ and achieve the maximal value at the
equatorial plane ($\theta = \pi/2$). On the contrary,
the two profiles of $c_s$ are significantly different.
In their Fig.~1, the value of $c_s$ decreases with increasing $\theta$
and achieves the minimal value at the equatorial plane; in our Fig.~1,
however, $c_s$ increases with increasing $\theta$
and achieves the maximal value at the equatorial plane. 
In our opinion, the difference results from different assumptions, i.e.,
NY95 assumed an energy advection factor $\fpadv$ in advance,
whereas we solve for the energy advection factor $\fadv$ self-consistently
based on a polytropic relation in the vertical direction.
We think that our profile for $c_s$ is reasonable for disk-like accretion.
For example, in the standard thin disk, 
the direction of the radiative flux is from the
equatorial plane to the surface, which means that the temperature
(or the sound speed) decreases from the equatorial plane to the surface.
Such a picture agrees with our Fig.~1 but conflicts with Fig.~1 of NY95.

Figure 2 shows the variation of $\fadv$ with $\Dtheta$ for the ratio
of specific heats $\gamma = 4/3$. Advection dominance means
$0.5 < \fadv \le 1$. We first explain the two dashed lines and
the dotted line that correspond to previous works in the slim disk model,
then the solid line that represents our results here, and leave
the dot-dashed line later. Both the two dashed lines are obtained
by assuming vertical hydrostatic equilibrium (eq. [3]) and using
the H\={o}shi form of potential (eq. [2]), thus the relation
$H \OmK /c_s =$ constant is adopted. The difference between these
two lines is the following. For line $a$, the simple one-zone treatment
in the vertical direction is made as in the SSD model;
then in equation (3),
$\p p/\p z \approx -p/H$, $\p \psi /\p z \approx \OmK^2 H$,
and $H \OmK/c_s = 1$ is obtained (e.g., Kato et al. 1998, p.80).
For line $b$, there is some improvement in the sense that the 
vertical structure of the disk is considered.
By assuming a polytropic relation, the vertical integration of
equation (3) gives $H \OmK/c_s = 3$ (e.g., Kato et al. 1998, p.242).
Because of these different treatments in the vertical direction,
these two lines show different variations of $\fadv$ 
with $\Dtheta$ and different maximum values of $\Dtheta$.
The upper limit of $\fadv$ is 1 (full advection dominance),
beyond which there would be no thermal equilibrium solutions.
It can be analytically derived that for the case of line $a$,
the maximum value of $\Dtheta$ corresponding to $\fadv = 1$ is
$\Dtheta_{\rm max} = \arctan(\sqrt{2/7})$,
or in cylindrical coordinates the maximum relative thickness
$(H/R)_{\rm max} = \sqrt{2/7}$; and for the case of line $b$
it is $\Dtheta_{\rm max} = \arctan(3/2)$ or $(H/R)_{\rm max} = 3/2$.
As mentioned in \S 1, the thickness of the disk in the slim disk model
had been underestimated because the vertical gravitational force
was overestimated by the H\={o}shi form of potential.
Even so, according to the more sophisticated version of the slim disk
model (line $b$), advection dominance $\fadv > 0.5$ would require
$H/R > 1$ ($\Dtheta > \pi/4$), and full advection dominance would
require $H/R = 3/2$, in contradiction with $H/R \lesssim 1$,
the supposed feature of the model.

The dotted line in Figure 2 is for the results of GL07.
The point made in that work was that the explicit potential $\psi (R,z)$,
rather than its H\={o}shi approximation (eq. [2]), was used,
so that the vertical gravitational force was correctly calculated.
But GL07 still kept the assumption of vertical hydrostatic equilibrium
(eq. [3]), i.e., the terms containing $v_z$ in equation (4)
were incorrectly ignored. Because of this, the thickness of the disk
was overestimated; and accordingly, it seemed that advection dominance
can never be possible, since even for the extreme thickness
$\Dtheta = \pi/2$ (or $H/R \to \infty$) the value of $\fadv$
can only marginally reach to 0.5.

We make improvements over GL07. We use spherical coordinates
with the assumption $v_{\theta} = 0$, which is better than $v_z = 0$
in cylindrical coordinates;
and then calculate the vertical distribution of velocities
($v_r$ and $v_{\phi}$) and thermal quantities ($\rho$, $p$, and $c_s$).
Our results are shown by the solid line in Figure 2. It is seen
that advection dominance ($\fadv > 0.5$) is possible,
but only for $\Dtheta > 2\pi /5$ (or 72$^{\circ}$).
Therefore, advection-dominated disks must be geometrically thick,
rather than slim as previously supposed.

It is also seen that line $b$, the dotted line, and the solid line
in Figure 2 almost coincide with each other for thin disks
with $\Dtheta \lesssim 0.1\pi$. This is natural, since for thin disks
both the H\={o}shi approximation of potential and the assumption
of vertical hydrostatic equilibrium are valid, and the three
approaches represented by the three lines make no significant difference.
But the one-zone treatment, i.e., total ignorance of
the vertical structure of the disk, seems to be too crude,
making the resulting line $a$ deviate from the other three lines
even for thin disks.

The value $\gamma = 4/3$ in Figure 2 corresponds to the optically
thick and radiation pressure-dominated case,
to which the historical radiation pressure-supported thick disk
and the slim disk belong; while it is $\gamma \to 5/3$ for the
optically thin and gas pressure-dominated case, to which the historical
ion pressure-supported thick disk and the ADAF belong.
In Figure 3, the four solid lines show variations of $\Dtheta$ with
$\gamma$ for four given values of $\fadv$. It is seen that
advection dominance ($\fadv > 0.5$) requires $\Dtheta$ to be large
for any value of $\gamma$; and that for a fixed $\fadv$
(the same degree of advection), the required $\Dtheta$ increases
with increasing $\gamma$, that is, for advection to be dominant,
optically thin disks must get even geometrically thicker
than optically thick ones.

For the geometrically thin case, $\Dtheta \ll 1$,
the Taylor expansion of equations (9), (11), and (12) with respect to
$\Dtheta$ can be performed, and we derive an approximate analytic relation:
\beq
\fadv \approx
\frac{(5-3\gamma)(2\gamma -1)}{3\gamma(5\gamma -3)} \cdot \Dtheta ^2 \ ,
\eeq
which is similar to equation (1) in cylindrical coordinates.
The dot-dashed lines in Figures 2 and 3 correspond to equation (17)
for a fixed $\gamma = 4/3$ and for a fixed $\fadv = 0.01$, respectively.
It is seen from Figure 2 that, as expected, the analytic approximation
of equation (17) agrees well with the correct numerical results
(the solid line) for small $\Dtheta$, but deviates a lot for
large $\Dtheta$. In Figure 3 a good agreement between equation (17)
and the numerical results (the lowest solid line) is seen again,
especially for small values of $\gamma$.
The limitation that equation (17) is valid only for small $\Dtheta$,
and accordingly only for small $\fadv$, should also apply to
equation (1), because that equation is derived with the H\={o}shi form
of potential.

\section{Discussion}

The key concept of the slim and ADAF disk models is advection dominance.
This concept was introduced rather as an assumption, whether and under what
physical conditions can it be realized have not been clarified.
The main result of our work is to have shown that, in order for advection
to be dominant, the disk must be geometrically thick with the half-opening
angle $\Dtheta > 2\pi /5$, rather than slim as suggested previously in the
slim disk and ADAF models.
Thus, advection-dominated disks are geometrically similar to the historical
thick disks metioned in \S 1.
This result is obvious because, as revealed
in GL07, in the slim disk and ADAF models the vertical gravitational force
was overestimated by using the H\={o}shi's approximate potential, and
accordingly the disk's thickness was underestimated.
NY95 considered accretion flows with no free surfaces and found that when
the given advective factor $\fpadv (\equiv \qadv / \qvis) \to 1$
(full advection dominance), their solutions approach nearly spherical
accretion. If ``nearly spherical" can be regarded as extremely thick,
then their results and ours agree with each other, but we take a different
approach. We do not give the value of $\fadv (\equiv \Qadv / \Qvis)$
in advance, but instead consider accretion flows with free surfaces,
i.e., accretion disks. The boundary condition is set to be $p = 0$,
which is usually adopted in the literature (e.g., Kato et al. 1998).
Then the thickness of the disk, $\Dtheta$, makes sense, and we calculate
$\fadv$ to see how it relates to $\Dtheta$.

Many 2D and 3D numerical simulations of viscous radiatively inefficient
accretion flows (RIAFs) revealed the existence of convection-dominated
accretion flows (CDAFs), while ADAFs could not be obtained
(e.g., Stone et al. 1999; Igumenshchev \& Abramowicz 2000;
McKinney \& Gammie 2002; Igumenshchev et al. 2003).
We think that this fact probably indicates that
the existing analytic ADAF models
might have hidden inconsistencies, and the incorrect treatment of the
vertical structure might be one such inconsistency,
as addressed in our work.
Moreover, the recent radiation-MHD
simulations (Ohsuga et al. 2009) showed that the disk is geometrically
thick in their models A and C (corresponding to slim disks and ADAFs,
respectively), which is in agreement with our results.

Apart from the convective motion, the outflow is found in 2D and 3D
MHD simulations of non-radiative accretion flows (e.g., Stone \& Pringle
2001; Hawley \& Balbus 2002). For optically thick flows, the circular
motion and the outflow are found in 2D radiation-HD simulations
(e.g., Ohsuga et al. 2005; Ohsuga 2006). The assumption $v_{\theta} = 0$
would break down when the convective motion or the outflowing motion is
significant, thus we have to point out the limitation of our solutions,
which are based on the self-similar assumption in the radial direction and
particularly for $v_{\theta} = 0$.

In this paper we have not shown the exact thermal equilibrium solution
for a certain mass accretion rate.
We wish to stress that our main concern here is the relationship between
the energy advection factor and the thickness of the disk.
The well-known formula (1), which was previously believed to be valid
for both optically thick and thin disks, implied that
advection-dominated accretion disks are geometrically slim.
As shown in Figures 2 and 3, however, formula (1) is inaccurate
for disks that are not geometrically thin.
We think that the new relationship between $\fadv$ and $\Dtheta$,
shown in Figures 2 and 3,
should also work for both optically thick and thin cases. Even without the
exact solutions, we can predict that advection-dominated accretion disks
ought to be geometrically thick rather than slim.
Our next work will concentrate on the optically
thick disks and take the radiative cooling into consideration.
In the vertical direction, we will solve the
dynamical equations combined with the radiative transfer equations,
thus the polytropic assumption will be relaxed.
At that step, we will be able to calculate the thermal equilibrium
solutions with given mass accretion rates and show the optical depth, pressure,
and luminosity of the disks.

\bigskip

We thank Marek A. Abramowicz, Ramesh Narayan, and Ken Ohsuga
for beneficial discussions and the referee for helpful comments.
This work was supported by the National Basic Research Program of China
under Grant No. 2009CB824800, the National Natural Science Foundation of
China under Grants No. 10778711 and 10833002, the Program
for New Century Excellent Talents in University under Grant No. 06-0559,
and the China Postdoctoral Science Foundation funded project 20080441038.

\newpage

\begin{figure*}
  \begin{center}
  \FigureFile(80mm,55mm){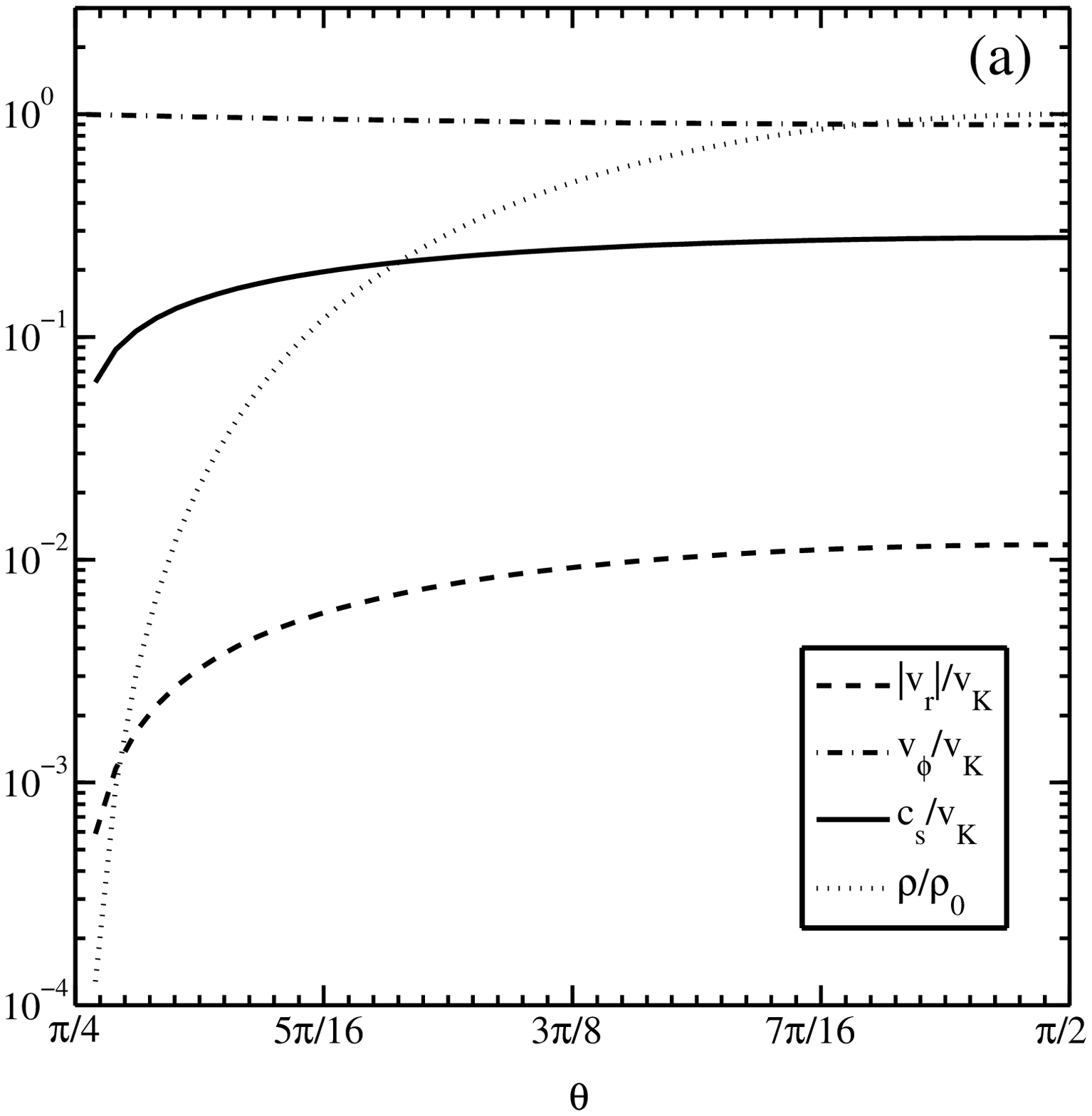}
  \FigureFile(80mm,55mm){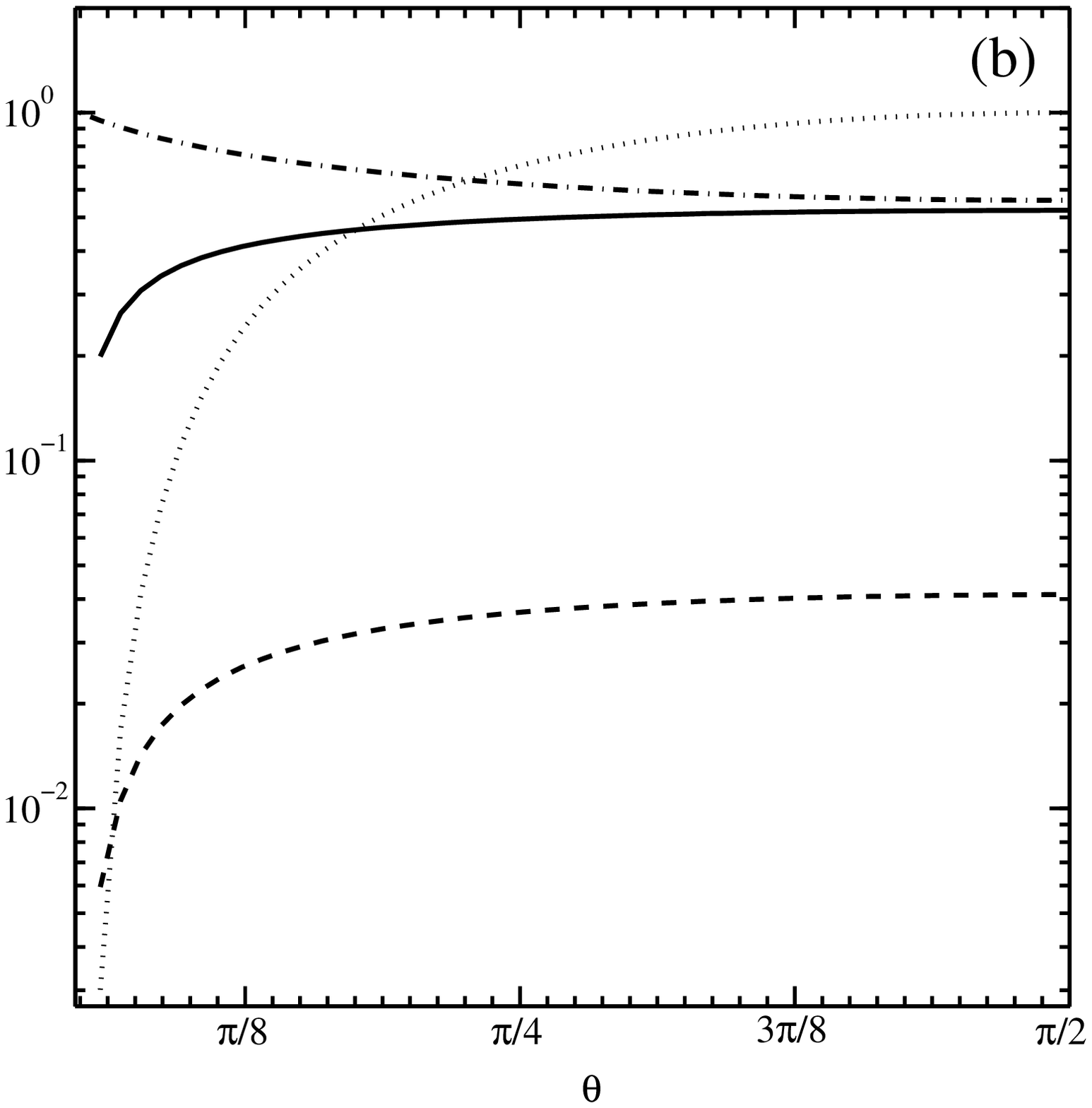}
  \FigureFile(80mm,55mm){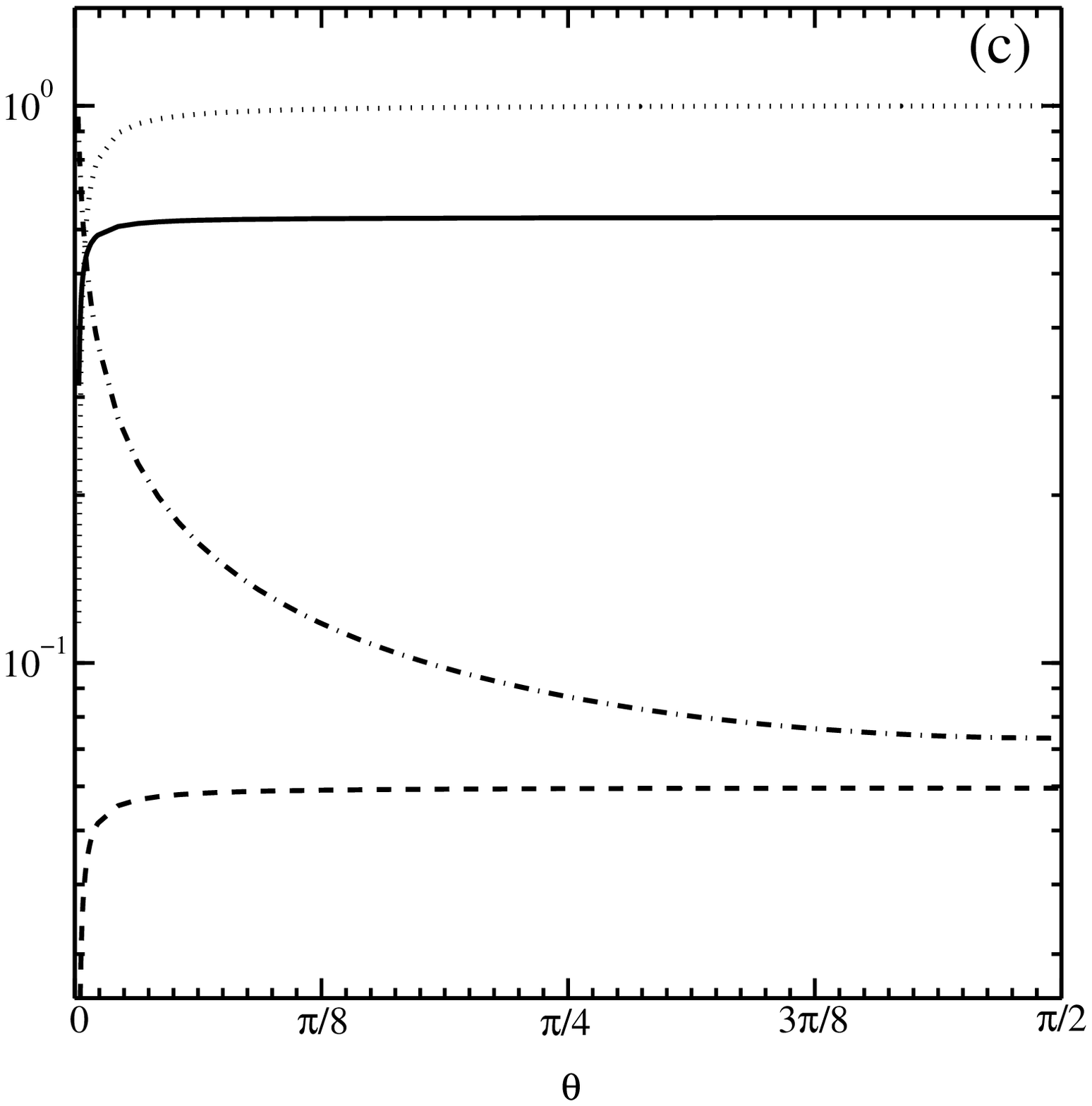}
  \end{center}
\caption{
Variations of $v_r$, $v_{\phi}$, $c_s$, and $\rho$ with the polar angle
$\theta$ for three pairs of parameters:
(a) $\gamma = 4/3$ and $\Dtheta = 0.25\pi$;
(b) $\gamma = 4/3$ and $\Dtheta = 0.45\pi$;
(c) $\gamma = 1.65$ and $\Dtheta = 0.498\pi$.
}
\label{fig1}
\end{figure*}

\newpage

\begin{figure*}
  \begin{center}
  \FigureFile(150mm,100mm){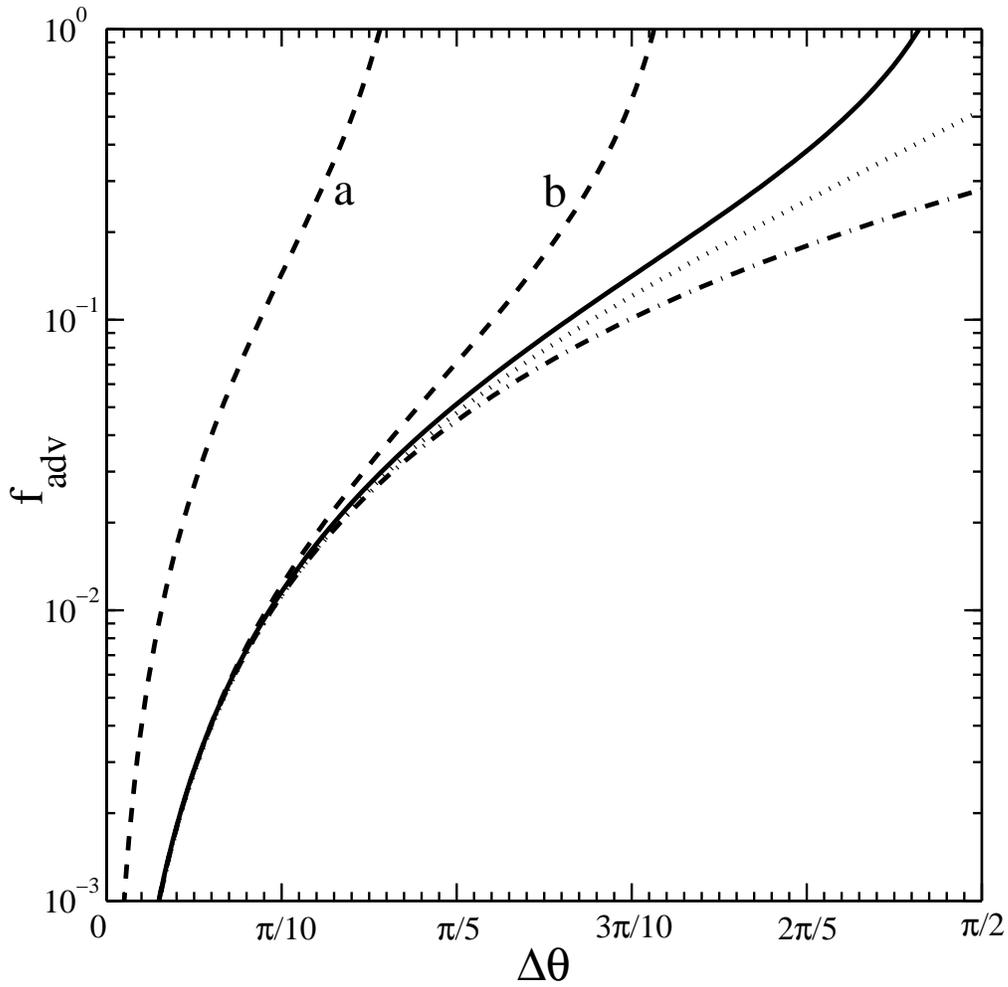}
  \end{center}
\caption{
Variation of the advection factor $\fadv$ with the disk's
half-opening angle $\Dtheta$ for the ratio of specific heats
$\gamma = 4/3$. The solid line shows our numerical results.
The dot-dashed line corresponds to the analytic approximation
of equation (17). The two dashed lines are for the previous results
in the slim disk model with the H\={o}shi form of potential,
and the dotted line is for the previous results in GL07.
}
\label{fig2}
\end{figure*}

\newpage

\begin{figure*}
  \begin{center}
  \FigureFile(150mm,100mm){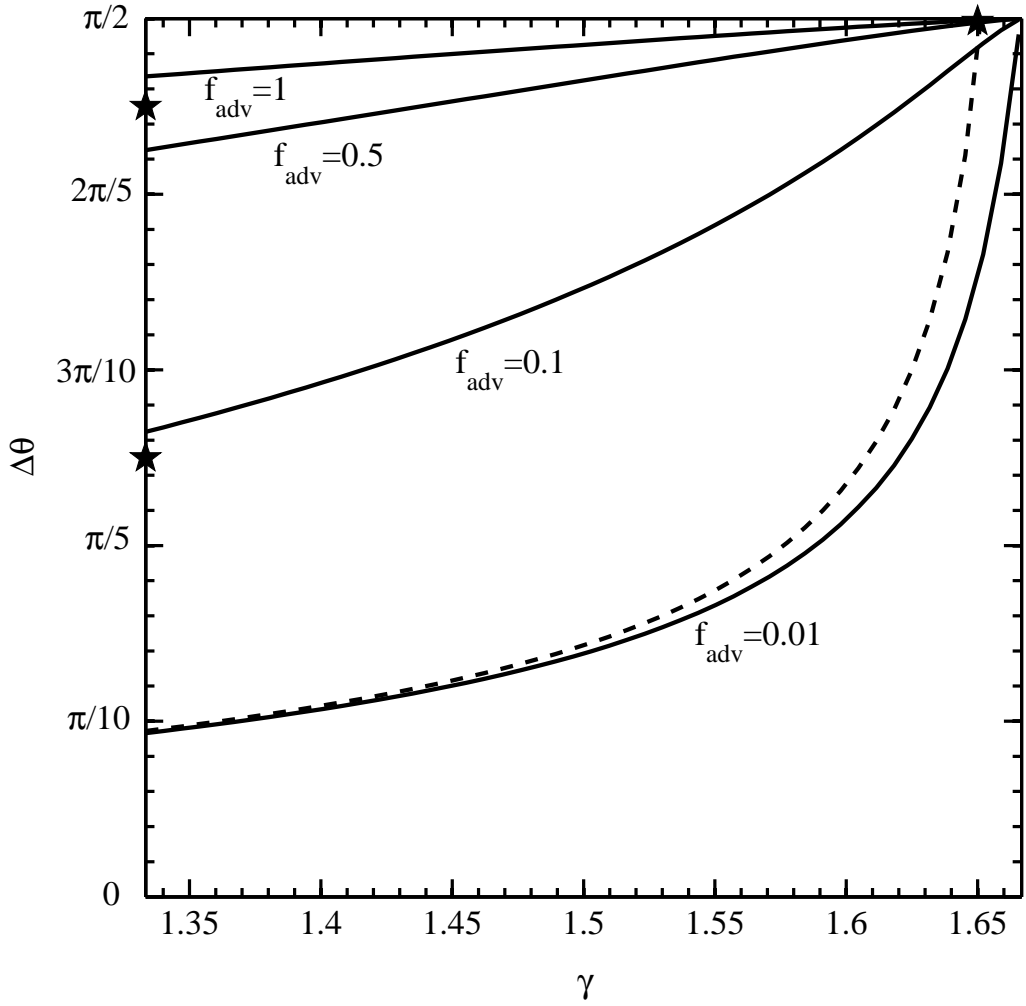}
  \end{center}
\caption{
Variation of $\Dtheta$ with $\gamma$ for given values of $\fadv$.
The solid lines show numerical results, and the dot-dashed line
corresponds to equation (17). The three filled stars denote the
parameters chosen in Figure~1.
}
\label{fig3}
\end{figure*}

\end{document}